\documentclass[aps,prd,twocolumn,amsmath,amssymb,showpacs,floatfix,nofootinbib]{revtex4-1}

\usepackage{graphicx}
\usepackage{dcolumn}
\usepackage{xcolor}
\usepackage{bm}
\usepackage{natbib}
\usepackage{calligra}
\usepackage[T1]{fontenc}
\usepackage{egothic}
\usepackage[T1]{fontenc}
\newfont{\rsfsten}{rsfs10 scaled 1200}
\newfont{\rsfsseven}{rsfs10 scaled 1200}
\newfont{\rsfsfive}{rsfs10 scaled 1200}
\usepackage{epsfig}
\usepackage{units}
\usepackage[utf8]{inputenc}

\newcommand{\be}{\begin{equation}}
\newcommand{\ee}{\end{equation}}
\newcommand{\bea}{\begin{eqnarray}}
\newcommand{\eea}{\end{eqnarray}}

\def\lsim{\mathrel{\raise.3ex\hbox{$<$\kern-.75em\lower1ex\hbox{$\sim$}}}}
\def\gsim{\mathrel{\raise.3ex\hbox{$>$\kern-.75em\lower1ex\hbox{$\sim$}}}}

\begin{document}

\vskip 0.2in

\title{On the Gravitational Wave Background from Black Hole Binaries after the First LIGO Detections}
\author{Ilias Cholis}
\email{icholis1@jhu.edu}
\affiliation{Department of Physics and Astronomy, The Johns Hopkins University, Baltimore, Maryland, 21218, USA}

\date{\today}

\begin{abstract}

The detection of gravitational waves from the merger of binary black holes by the LIGO Collaboration
has opened a new window to astrophysics. 
With the sensitivities of ground based detectors in the coming years, we
will principally detect local binary black hole mergers.
The integrated merger rate can instead be probed by the gravitational-wave background, the incoherent superposition of the released energy in gravitational waves during binary-black-hole coalescence.
Through that, the properties of the binary black holes 
can be studied. In this work we show that by measuring the energy density $\Omega_{GW}$ (in units of the cosmic critical density) of the gravitational-wave background, we can search for the rare $\sim 100 M_{\odot}$
 massive black holes formed in the Universe. In addition, we can answer how often the least massive BHs of 
mass $\gsim 3 M_{\odot}$ form. Finally, if there are multiple channels for the formation of binary black
holes and if any of them predicts a narrow mass range for the black holes, then the total $\Omega_{GW}$
spectrum may have features that with the future Einstein Telescope can be detected.
\end{abstract}


\maketitle

\section{Introduction}
\label{sec:introduction}

The observation of the coalescence of black holes (BHs) by the LIGO collaboration~\cite{Abbott:2016blz, Abbott:2016nmj, TheLIGOScientific:2016pea}, has generated great interest in gravitational wave (GW) physics and in the sources
responsible for them. Many alternatives have been proposed regarding the progenitors of binary black holes (BBHs), including from BH as the end product of stellar evolution of massive stars \cite{Hosokawa:2015ena, Zhang:2016rli, Woosley:2016nnw, deMink:2016vkw, Hartwig:2016nde, Inayoshi:2016hco}, of BHs in globular clusters \cite{Rodriguez:2016kxx, Chatterjee:2016hxc, Rodriguez:2016avt, O'Leary:2016qkt}, or in centers of galaxies \cite{O'Leary:2008xt, Stone:2016wzz}, or as the result of primordial black holes capturing each other \cite{Bird:2016dcv, Clesse:2016vqa, Sasaki:2016jop}, all consistent with the observed BBH merger rates \cite{Abbott:2016nhf, TheLIGOScientific:2016pea}. In order to probe and discriminate among the various models for the 
BBH progenitors, different observables will be necessary. With LIGO, we expect that BH binaries of 
composite masses of 10 (20, 30) $M_{\odot}$ will be detectable as individual events, only up to redshifts of 0.3 (0.5, 0.7). However, with the future Einstein Telescope (ET) \cite{Sathyaprakash:2011bh} those redshifts may increase up to 11 (12, 11) respectively. 

In addition, the entire merging BBH population of the Universe, can be probed through the incoherent superposition of their released energy in GWs, giving the gravitational wave background \cite{Kosenko:1998mv, Ferrari:1998ut, Schneider:2000sg, Schneider:2000sg, Hogan:2001jn, Farmer:2003pa, Howell:2010hm, Regimbau:2011rp, Rosado:2011kv, Zhu:2011bd, Marassi:2011si, Wu:2013xfa, Regimbau:2014uia, TheLIGOScientific:2016wyq}. This background is affected, by both the BBH population mass and redshift  distributions at the time of the merger. In turn these distributions depend on the environment where the BH binaries form. 

After its first run, advanced LIGO has detected two events, event GW150914 of $36.2^{+5.2}_{-3.8}$ and $29.1^{+3.7}_{-4.4}$ $M_{\odot}$ merging BHs, and event GW151226 of $14.2^{+8.3}_{-3.7}$ and $7.5^{+2.3}_{-2.3}$ $M_{\odot}$
, each with a significance larger than 5.3 $\sigma$. LIGO has also detected one possible event, LVT151012 of $23^{+18}_{-6}$ and $13^{+4}_{-5}$ $M_{\odot}$ with a significance of 1.7 $\sigma$ \cite{TheLIGOScientific:2016pea}. 
Merging BHs, during the last stages of their coalescence when most energy is radiated, emit GWs at frequencies and with amplitudes that depend on the 
combination of their masses.
Using the first three events, and the estimated BBH merging rates, we can study the impact that uncertainties on individual BH mass and redshift distributions have on the gravitational wave background, and present how those properties can be further probed. 

This paper is organized as follows; in section~\ref{sec:Methodology} we give the basic set up for our calculations and in section~\ref{sec:PBHs_OmegaGW} we give our main results and discuss on the detectability of the gravitational wave background. Finally in section~\ref{sec:Conclusions} we give our conclusions. 
               
\section{The Gravitational Wave background from Binary Black Holes}
\label{sec:Methodology}

The total energy density spectrum of gravitational waves is given by:
\begin{equation}
\Omega_{GW}(f_{\textrm{obs}}) = \frac{1}{\rho_{c}}\frac{d\rho_{GW}}{d ln f_{\textrm{obs}}},
\label{eq:OmegaGWdef}
\end{equation}
where $f_{\textrm{obs}}$ is the observable GW frequency, $d\rho_{GW}$ is the GW energy density between $f_{obs}$ and $f_{obs}+df_{obs}$ and 
$\rho_{c}$ is the critical energy density of the Universe $\rho_{c} = 3 H_{0}^2/(8 \pi G)$.
There can be various other contributions to the $\Omega_{GW}$ (see e.g. \cite{Gasperini:1992dp, Ferrari:1998jf, Farmer:2003pa, Regimbau:2005ey, Carbone:2005nm, Smith:2005mm,  Ananda:2006af, Siemens:2006yp, GarciaBellido:2007af, Caprini:2009yp, Olmez:2010bi, Binetruy:2012ze, Arzoumanian:2015liz, Domcke:2016bkh, Dvorkin:2016wac, TheNANOGrav:2016ihr}).
From this point on we are going to study only the contribution from BH binaries, since for those we 
already have observations and thus some first measurement of their local incidence rate.
Accounting for the fact that the GWs are emitted from the coalescing binaries with a spectral energy density $dE/df$, we have for the total, resolved and unresolved population of coalescing binaries:
\begin{equation}
\Omega_{GW}(f_{\textrm{obs}}) = \frac{f_{\textrm{obs}}}{c^{2} \rho_{c}} \int_{0}^{z_{\textrm{max}}} dz \frac{R_{m}(z)}{(1+z)H(z)}\frac{dE}{df_{s}},
\label{eq:OmegaGW}
\end{equation}
where $z_{\textrm{max}}$ is the maximum redshift relevant for the sources of the GWs and for the frequencies sensitive to the observatories. $R_{m}(z)$ is the merger rate of the BBHs at their source and $f_{s}$ is the GW frequency at the source ($f_{s} = f_{\textrm{obs}}(1+z)$). $H(z) = H_{0} \sqrt{\Omega_{M}(1+z)^{3}+\Omega_{\Lambda}}$.

The energy released during the inspiral, the merger and the ring-down phases of the coalescence and the exact frequencies of GWs emitted, has been the question of extended studies \cite{Buonanno:2002fy, Blanchet:2004ek, Blanchet:2006gy, Blanchet:2008je, Arun:2008kb, Ajith:2009bn, Ajith:2011ec}.
To account for the uncertainties related to the energy density released during the coalescence, we follow two alternative parametrizations of the spectral energy density of the emitted GWs. The first we refer to as Ajith et al. \cite{Ajith:2009bn} and the second is referred to as
Flanagan $\&$ Hughes \cite{Flanagan:1997sx}.

Both parametrizations agree on the spectral energy density of the emitted GWs during the inspiral of a circularized orbit, which at the source is:
\begin{eqnarray}
\frac{dE}{df_{s}}_{\textrm{insp}} = \frac{1}{3} \left( \frac{\pi^{2} G^{2}}{f_{s}} \right)^{1/3} \frac{m_{2} \cdot m_{2}}{(m_{1}+m_{2})^{1/3}}.
\label{eq:dEdf_Insp}
\end{eqnarray}
The frequency at the end of the inspiral and the beginning of the merger phase is (at the source):
\begin{eqnarray}
f_{\textrm{merg}} (m_{1}, m_{2}) = 0.02 \frac{c^{3}}{G(m_{1}+m_{2})},
\label{eq:fmerg}
\end{eqnarray}
accourding to \cite{Flanagan:1997sx} and double that value for BBHs according to \cite{Ajith:2009bn}. 
Between the redshifted $f_{\textrm{merg}}$ and the frequency of quasi-normal ring-down (at the position of the binary) \cite{Flanagan:1997sx}:
\begin{eqnarray}
f_{\textrm{qnr}} (m_{1}, m_{2}) = \frac{c^{3} \left(1-0.63(1-\alpha)^{3/10} \right)}{2 \pi G (m_{1}+m_{2})},
\label{eq:fqnr}
\end{eqnarray}
the merger phase of the coalescence event is observed.
$\alpha$ is the dimensionless spin of the final BH; $\alpha = \frac{c S}{G m_{\textrm{final}}^{2}}$, assuming $m_{\textrm{final}} \simeq m_{\textrm{tot}} = m_{1}+m_{2}$.
 The quasi-normal ring-down frequency $f_{\textrm{qnr}}$ by \cite{Ajith:2009bn} is only 8$\%$ less than that of \cite{Flanagan:1997sx} for a given choice of masses and spins. More recent models
 \cite{Ajith:2007qp, Santamaria:2010yb, Khan:2015jqa}, provide even more accurate expressions on the amplitude of the GWs versus masses and spins.
We take $\alpha = 0.67$ thought this paper, given that the measured values from the first observations indicate such a value. We also note that its choice has a minimal impact in the calculations of the total
released energy density in GWs for coalescence events.

During the merger phase, the spectral energy density is given in turn by \cite{Flanagan:1997sx}:
\begin{eqnarray}
\frac{dE}{df_{s}}_{\textrm{merger}} = \frac{16 c^{2} \mu^{2} \epsilon}{m_{\textrm{tot}}(f_{\textrm{qnr}}(m_{1},m_{2}) - f_{\textrm{merg}}(m_{1}, m_{2}))},
\label{eq:dEdf_Merg}
\end{eqnarray}
where $\mu$ is the reduced mass and $\epsilon$ is the fraction of the energy in the initial BH binary that is emitted in GWs during that phase. We take $\epsilon = 0.04$ in agreement with the uncertainties of the GW150914 event \cite{TheLIGOScientific:2016wfe}. 
Following the parametrization of \cite{Ajith:2009bn}, we get instead: 
\begin{eqnarray}
\frac{dE}{df_{s}}_{\textrm{merger}} = \frac{1}{3} \left(\pi^{2} G^{2}\right)^{1/3} \frac{f_{s}^{2/3}}{f_{\textrm{merg}}}  \frac{m_{2} \cdot m_{2}}{(m_{1}+m_{2})^{1/3}}.
\label{eq:dEdf_Merg_Ajith}
\end{eqnarray}

Finally, the energy density during the ring-down phase, in given by \cite{Flanagan:1997sx} as:
\begin{eqnarray}
\frac{dE}{df_{s}}_{\textrm{qnr}} = \frac{G^{2}A^{2}Q}{8}m_{\textrm{tot}}^{2} f_{\textrm{qnr}},
\label{eq:dEdf_QNR}
\end{eqnarray}
with $A=0.4$, $Q = 2(1-\alpha)^{-9/20}$ and by \cite{Ajith:2009bn} as:
\begin{eqnarray}
\frac{dE}{df_{s}}_{\textrm{qnr}}&=& \frac{1}{3} \left(\pi^{2} G^{2}\right)^{1/3} \frac{1}{f_{\textrm{merg}} f_{\textrm{qnr}}^{4/3}} \left( \frac{f_{s}}{1+ 4\left(\frac{f_{s}-f_{\textrm{qnr}}}{\sigma} \right)^2} \right)^2 \nonumber \\ && \frac{m_{2} \cdot m_{2}}{(m_{1}+m_{2})^{1/3}},
\label{eq:dEdf_QNR_Ajith}
\end{eqnarray}
with $\sigma = 237(20 M_{\odot}/m_{\textrm{tot}})$ Hz.
In either parametrization, the energy released during the quasi-normal ringdown is subdominant.

The second astrophysics input to the $\Omega_{GW}$ calculation, is
the merger rate of BBHs, $R_{m}(z)$. This is given by the convolution of the rate of binary formation $R_{f}(z)$ and the time delay $t_{d}$ distribution $P(t_{d})$ that describes the time it takes for those binaries to merge,
\begin{eqnarray}
R_{m}(z) = \int_{t_{\textrm{min}}}^{t_{H}} R_{f}(z_{f})P(t_{d}) dt_{d}.
\label{eq:Rm}
\end{eqnarray}
$t_{\textrm{min}}$ is the model dependent minimum value for $t_{d}$, $t_{H}$ the Hubble time and $z_{f}$ is the redshift at the formation.

\begin{figure}
\begin{centering}
\includegraphics[width=\columnwidth]{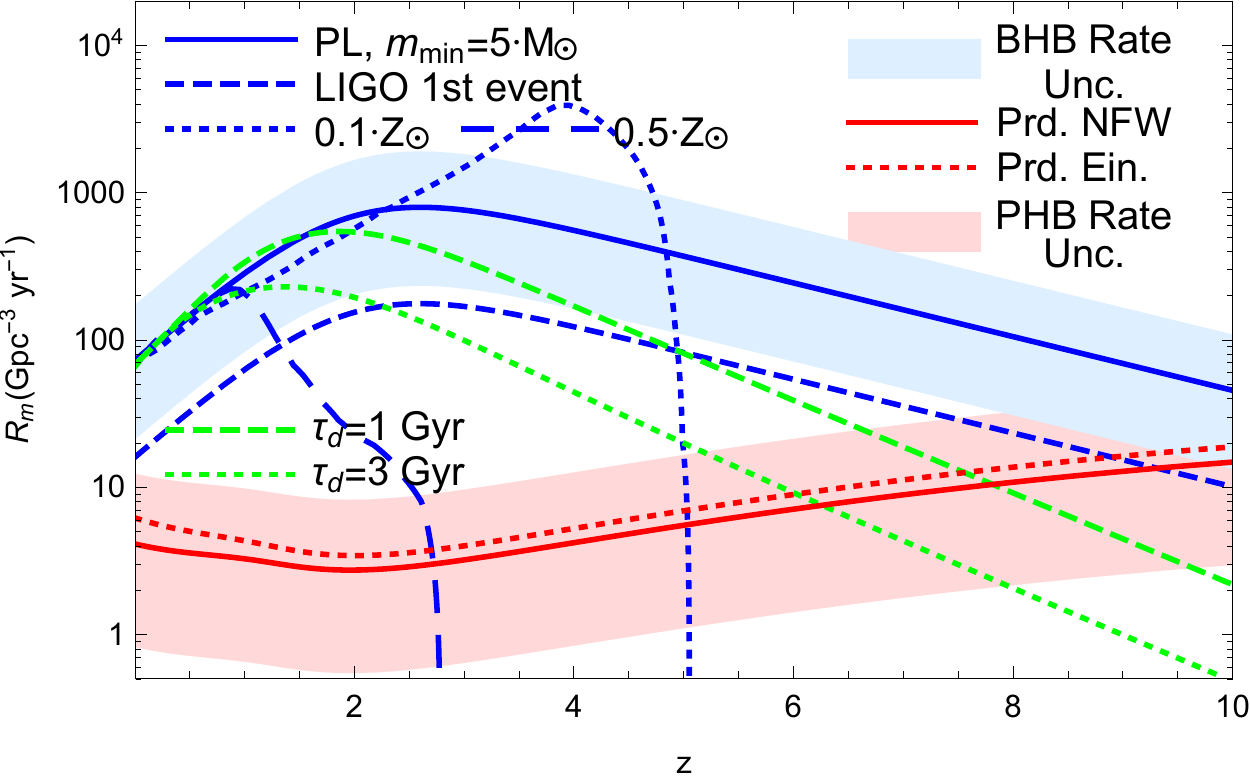}
\end{centering}
\caption{The comoving rate of binary BH mergers. The blue dashed line gives the LIGO fiducial assumption for the local rate 
\cite{TheLIGOScientific:2016wyq}, with the star formation rate by \cite{Vangioni:2014axa}, while the blue solid line and band give for the same redshift dependence the updated local merger rate of $R_{m} = 99^{+138}_{-70}$ Gpc$^{-3}$yr$^{-1}$ from \cite{TheLIGOScientific:2016pea} (see text for details). Given the lack of knowledge on the BBHs progenitors, 
the redshift dependence in $R_{m}(z)$ can vary significantly, as is shown by the dotted and long-dashed blue lines (normalized to the updated rate), where the BBH progenitors come from environments with metallicity of 0.1 and 0.5 $Z_{\odot}$ respectively \cite{Dominik:2013tma}. 
All blue lines assume no significant time delay between the formation and the merger of the binary. Instead, for the green dashed and dotted lines, a time delay of 1 and 3 Gyrs is assumed; with a binary formation rate following \cite{Vangioni:2014axa}, and normalized to the rate of \cite{TheLIGOScientific:2016pea}.  
As an example the PBH binaries also have large uncertainties due to the DM profile (solid versus dotted red lines),
the DM mass-concentration relation (shown for the one of Prada \cite{Prada:2011jf}) and the exact contribution of the smallest in mass DM halos, represented by a red band around the Prada NFW red solid line (see \cite{Mandic:2016lcn} for a detailed discussion). }
\label{fig:MergerRates}
\end{figure}

With the first LIGO event, the merger rate for BBHs was measured to be between 2 and 53 Gpc$^{-3}$yr$^{-1}$ locally, assuming all events were of the GW150914 class \cite{Abbott:2016nhf}.
By "class", in this work we assume that there are many BBHs with the same properties (masses and spins) as those measured at the event defining that class.
In Figure~\ref{fig:MergerRates}, we use that merger rate to produce the blue dashed line, assuming
that BBHs are formed with a rate that follows the star formation rate (SFR), probed by the observations of gamma-ray bursts (GRB) \cite{Vangioni:2014axa}, and ignoring for simplicity the time-delay between the formation and the merger of the binaries.  
With the three events of LIGO's complete O1 run, $R_{m}(z)$ has changed to $99^{+138}_{-70}$ Gpc$^{-3}$yr$^{-1}$ \cite{TheLIGOScientific:2016pea}, assuming that the larger mass $m_{1}$ of the BHs in the binaries,
follows a mass function $\propto m_{1}^{-2.35}$ with $m \ge 5 M_{\odot}$, and $m_{1}+m_{2} \le 100 M_{\odot}$. With otherwise the same assumptions on time-delay and formation rate, the updated merger rate is shown with its uncertainty by the solid blue line and the blue band around it in Figure~\ref{fig:MergerRates}. It is clear that rate is yet very uncertain. 

Given that BHs come from massive stars that lie on the massive end of the initial stellar mass function, the metallicity $Z$, of the environment where BBHs form, can have a strong impact on their formation redshift distribution. The blue dotted and long dashed lines in Figure~\ref{fig:MergerRates}, show that impact
assuming that all the BBHs form in environments of $Z = 0.1 Z_{\odot}$ or $Z = 0.5 Z_{\odot}$ respectively. For the redshift distributions of different metallicity environments we follow the results of \cite{Dominik:2013tma}.
Realistically, BBHs will form in a variety of environments, with relative weights not well defined yet. As is 
evident from Eq.~\ref{eq:Rm}, the time-delay between formation and merger, can be important as well.
Many models for BBHs suggest a typical time-scale for $t_{d}$ of the order of Grys \cite{Kulkarni:1993fr, Kalogera:2003pk,
Kalogera:2006uj, Vanbeveren:2008sj, Dominik:2013tma,
Mennekens:2013dja, Dominik:2014yma, Mandel:2015qlu,
Chatterjee:2016hxc}. Those large time-scales naturally arrise form the orbital properties (eccentricities and semi-major axes) of the formed binaries.  We assume that the propability density function $P(t_{d}) \propto exp\{-t_{d}/\tau_{d}\}$. In  Figure~\ref{fig:MergerRates}, using the same assumptions as otherwise used for the solid blue, we show the impact of the time-delay, where with the green dashed line we take $\tau_{d} = 1$ Gyr and with the green dotted line $\tau_{d} = 3$ Gyrs. The LIGO collaboration also models the time delay
distribution as $\propto 1/t_{d}$. We show results later using that alternative parametrization. 

Finally, we show a case for an alternative scenario to that of BHs that are formed from stars. That is the case presented in \cite{Bird:2016dcv}, where the BHs are primordial in their origin. These primordial black holes (PBHs) binaries, follow a very different redshift profile at formation. The binaries are formed through GW emission when PBHs pass close-by and emit enough energy to capture each other. The time delay
for these binaries was shown in \cite{Cholis:2016kqi} to be significantly smaller than that in conventional stellar BBHs. Yet, the capture rate has significant uncertainties as well. Those are related to the dark matter (DM) 
halo profiles (see red solid vs dotted lines for NFW \cite{Navarro:1995iw} vs Einasto \cite{1965TrAlm...5...87E} DM profiles), 
the mass concentration relation as well as uncertainties related to the contribution of the smallest DM 
halos shown in Figure~\ref{fig:MergerRates} with a red band (see also \cite{Bird:2016dcv, Mandic:2016lcn} for a detailed discussion of these effects).

\section{Results. The gravitational wave energy density probing the binary black holes properties}
\label{sec:PBHs_OmegaGW}

\begin{figure}
\begin{centering}
\includegraphics[width=\columnwidth]{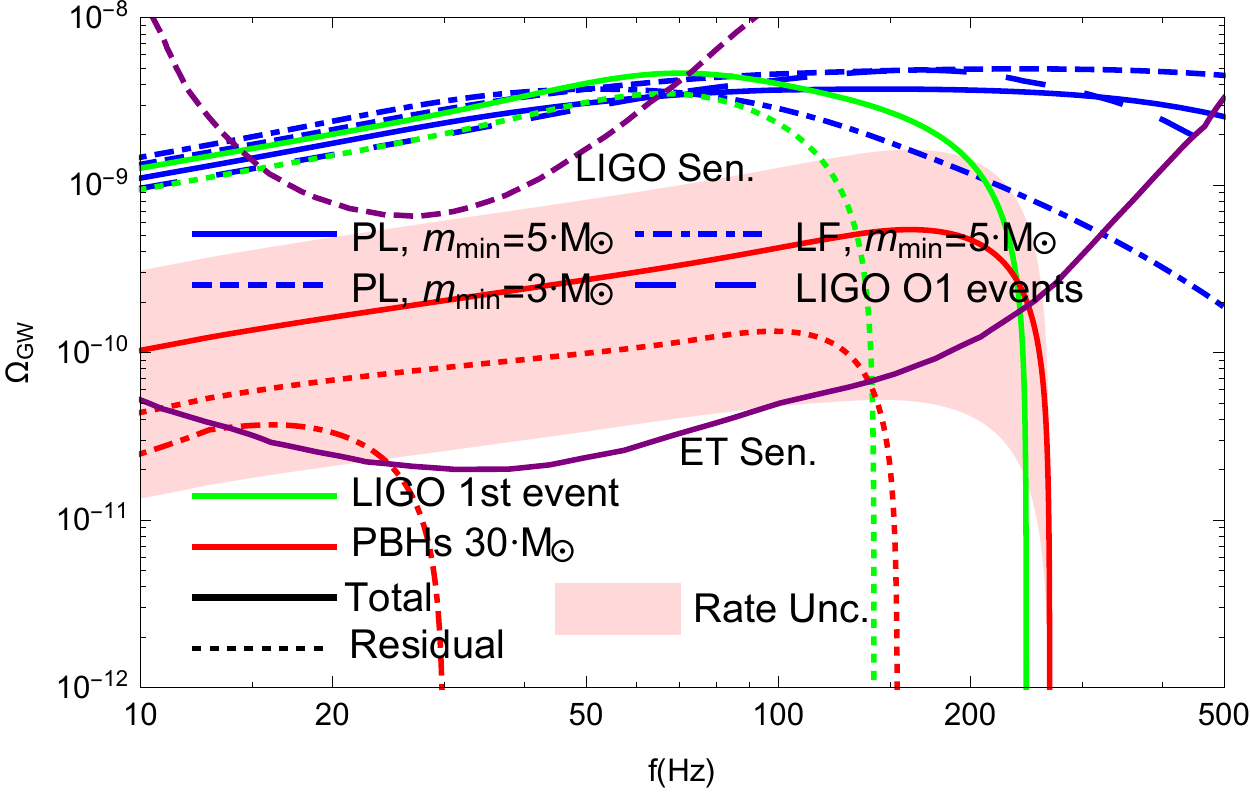}
\end{centering}
\caption{The gravitational wave energy density $\Omega_{\textrm{GW}}$, from the merger of stellar BBHs and of PBH binaries. 
The blue lines assume different mass distributions for the stellar BBHs with negligible time delay. For the solid, dashed and dot-dashed lines, the binary formation rate follows \cite{Vangioni:2014axa}, with the local rate in agreement with \cite{TheLIGOScientific:2016pea}.
The long-dashed blue line gives the $\Omega_{\textrm{GW}}$, assuming that the three events from 
LIGO O1 run are a representative sample of the entire population (see text for details and Figure \ref{fig:StochGWBackMassAssup}). The contribution of the PBHs, is given by the red lines for $m_{\textrm{PBH}} = 30 M_{\odot}$. The red band around the red solid line accounts for uncertainties in the rate (see text for more details and Figure \ref{fig:MergerRates}).
"Total" refers to the energy density calculated by the integral of Eq.~\ref{eq:OmegaGW}. As "Residual" we exclude the contribution from individual events that would be identified by LIGO's design sensitivity. We also show in the dashed-dotted red line the residual (stochastic) background from $30 M_{\odot}$ PBHs assuming that ET would resolve all coalescence events up to at least $z \simeq 8$. For comparison we also give the GW energy density calculated by LIGO  \cite{TheLIGOScientific:2016wyq} based on its first GW150914 event (green lines). The LIGO design and the ET expected sensitivities to the $\Omega_{\textrm{GW}}$ are given by purple dashed and solid lines respectively.}
\label{fig:StochGWBack}
\end{figure}

As we discussed in section \ref{sec:introduction}, many different sources contribute to the 
gravitational wave energy density $\Omega_{GW}$. Thus, with future detector 
upgrades, by measuring it in a wide range of frequencies, we will be able to probe the
properties of its sources. In Figure~\ref{fig:StochGWBack},  we show the gravitational waves energy density  
from BBHs using the estimated rate of BBH mergers, from the LIGO collaboration 
\cite{Abbott:2016nhf, TheLIGOScientific:2016pea}.
The green solid and dashed lines, give the estimated total and residual  GW energy density 
based on the first estimate by LIGO for the rate of events similar to the GW150914 \cite{Abbott:2016blz}. 
As residual we exclude from the total $\Omega_{GW}$, the contribution of all the mergers of that type, that would be identifiable as single events. For the composite masses of the GW150914 type, with the same spins and released energy in GWs, that is all mergers occurring within $z \le 0.75$. In the blue lines, we give instead the expected $\Omega_{GW}$, using the updated rate estimates. 

In \cite{TheLIGOScientific:2016pea}, the LIGO collaboration has presented three different estimates
for the rate of BBH mergers. Assuming that in the binaries, the most massive BH $m_{1}$, follows a 
probability distribution scaling as $\propto m_{1}^{-2.35}$, with the additional assumption that the 
mass ratio $q=\frac{m_{2}}{m_{1}}$, follows a flat distribution between a minimum mass of $5 M_{\odot}/m_{1}$ and 1. If instead the probability distribution for $m_{1}$ and $m_{2}$ scales 
as $p(m_{1}, m_{2}) \propto m_{1}^{-1} m_{2}^{-1}$, the local BBH merger rate is $30^{+43}_{-21}$ Gpc$^{-3}$yr$^{-1}$. That second assumption is referred to as flat in logarithmic mass (LF), with $m_{\textrm{min}}=5 M_{\odot}$. 
Finally, given the three measured events, LIGO has estimated the updated rates for the three classes 
of events to be $3.4^{+8.6}_{-2.8}$ Gpc$^{-3}$yr$^{-1}$ for the GW150914 class, $9.4^{+30.4}_{-8.7}$
Gpc$^{-3}$yr$^{-1}$ for the LVT151012 class and $37^{+92}_{-31}$ Gpc$^{-3}$yr$^{-1}$ for the GW151226 class. We refer to that as "LIGO O1" events assumption. We note that LIGO is less sensitive to masses $\lsim 5 M_{\odot}$. BHs with mass as low as 3 $M_{\odot}$ may exist in binaries even if 
observationally those haven't been detected in x-rays \cite{Kreidberg:2012ud}. 
Thus we allow also for the possibility that the distribution describing the largest masses $m_{1}$ in BBHs, extends down to
3 $M_{\odot}$ (with the ratio $q$ still following a flat distribution). As is shown 
in Figure~\ref{fig:StochGWBack} (blue solid, dot-dashed, long dashed and dashed lines) varying among those assumptions, affect little the lower frequencies $f \leq 50$Hz shown for the $\Omega_{GW}$.
Instead, at higher frequencies those varying assumptions, have a dramatic impact, since the lower masses occurrence, varies significantly between those alternative distributions. 

The total contribution at $20 < f_{\textrm{obs}} < 50$ Hz, will be probed by the LIGO final design sensitivity as is depicted by the purple
dashed line from \cite{TheLIGOScientific:2016wyq}.  
That assumes the BBHs merger rates are those quoted in the central values of \cite{TheLIGOScientific:2016pea}.
If instead the rate is lower but still within the currently quoted uncertainties, LIGO may not be able to
measure the $\Omega_{GW}$ at any frequency. As an example of that, in the red solid and dotted lines we give the contribution to the GW energy density from the PBH binaries. For $\sim$30 $M_{\odot}$ PBHs have
a merger rate that is consistent with the GW150914 class of events \cite{Bird:2016dcv}.
In the red band we give the uncertainty on $\Omega_{GW}$, based on the uncertainties on that rate
 (see Figure~\ref{fig:MergerRates} and also \cite{Mandic:2016lcn}).
Yet, with ET-B design \cite{Sathyaprakash:2012jk} (solid purple in Figure~\ref{fig:StochGWBack}), we expect that we will have enough 
sensitivity to measure $\Omega_{GW}$ from such a class of events over frequencies of 10-300 Hz; even for such a low merger rate. 
As a measure of the sensitivity to a given $\Omega_{GW}$ we use the signal to
noise ratio $S/N$, which is defined as:
 \begin{eqnarray}
S/N = \frac{4 G \rho_{c}}{5 \pi c^{2}} \sqrt{2 T} \left( \int_{f_{\textrm{min}}}^{f_{\textrm{max}}} df \frac{\Omega_{GW}^{2}(f) \gamma^{2}(f)}{f^{6}S_{n}^{2}(f)} \right)^{0.5},
\label{eq:StoN}
\end{eqnarray}
where $S_{n}(f)$ is the spectral noise density of each detector. We follow 
\cite{Moore:2014lga} \footnote{See also \url{http://rhcole.com/apps/GWplotter/}} to calculate the $S_{n}(f)$ from the plotted sensitivities in Figure~\ref{fig:StochGWBack}. $\gamma(f)$ is the dimensionless overlap reduction 
function for for ET we take it to be 1 for simplicity. For LIGO (ET) $f_{\textrm{min}} = 10 (1)$ Hz and $f_{\textrm{min}} = 1000$ Hz. For the total observation time we assume 5 yrs. 

\begin{figure}
\begin{centering}
\includegraphics[width=\columnwidth]{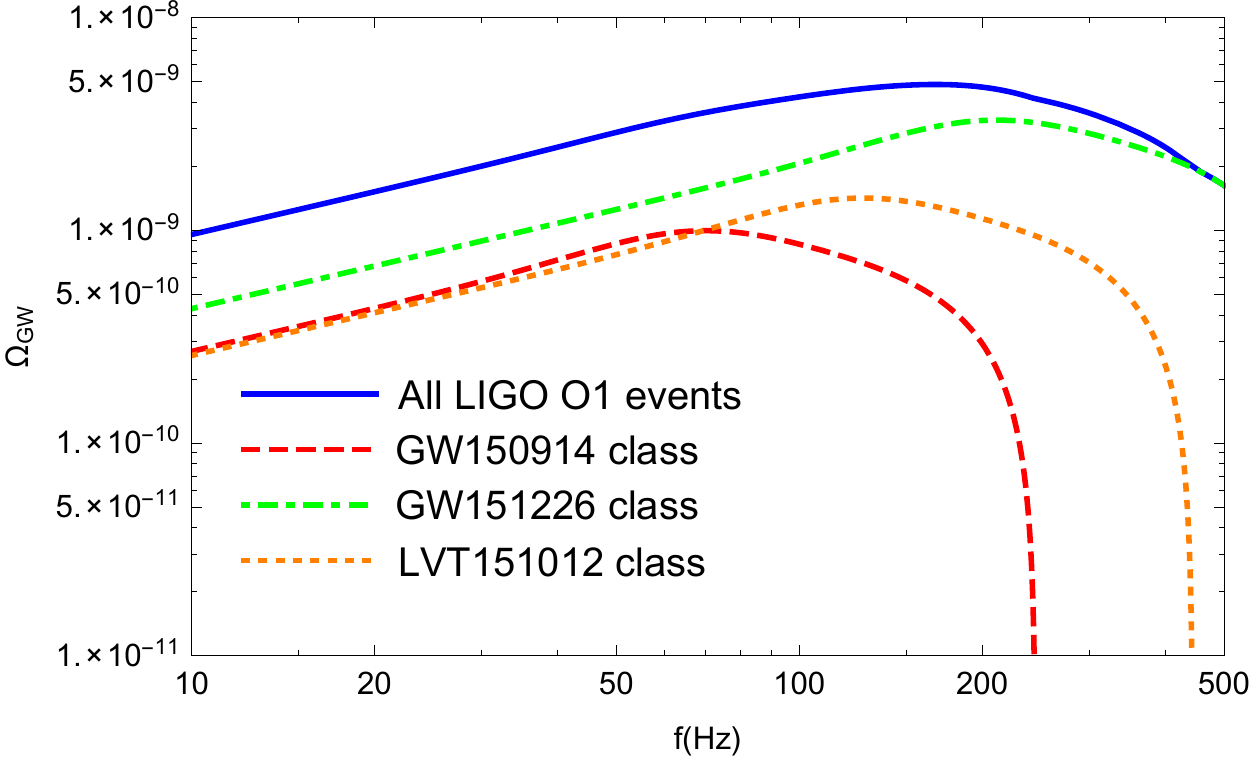}\\
\includegraphics[width=\columnwidth]{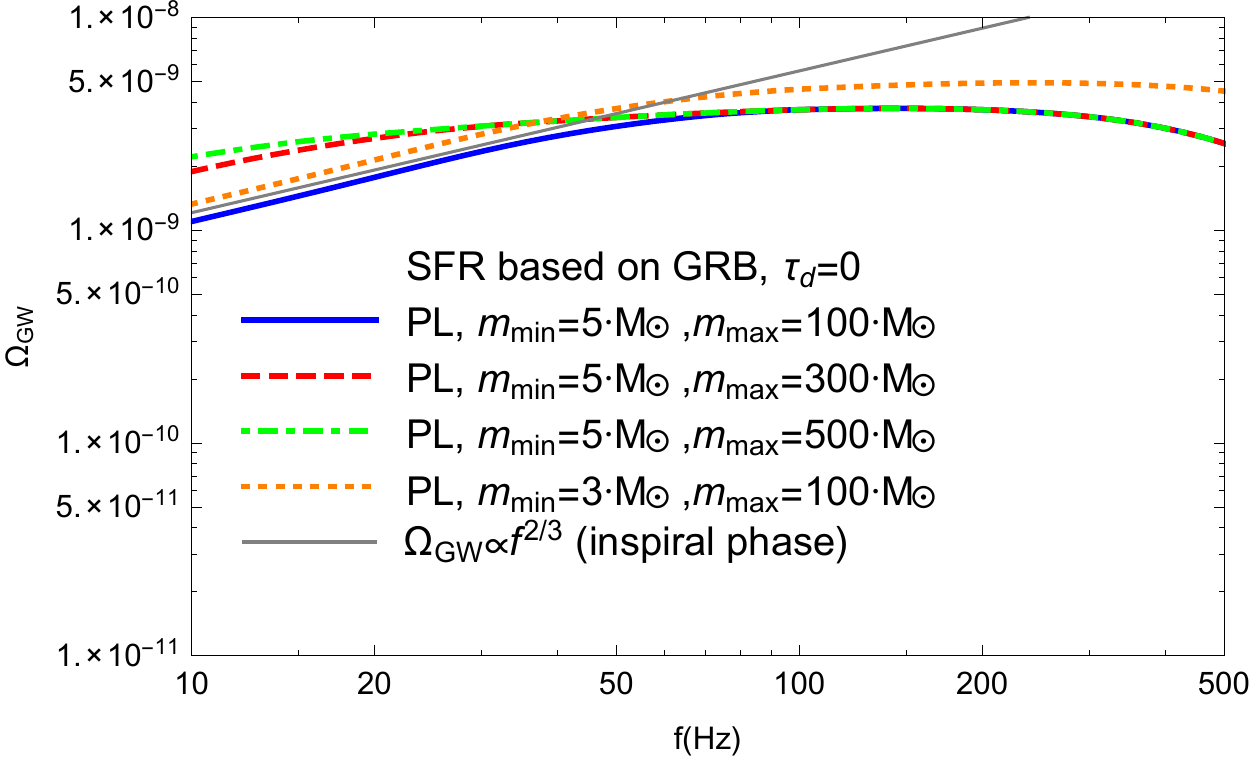}
\end{centering}
\caption{\textit{Top}: assuming that the three events observed by LIGO O1 run give a representative sample of the population of the BBHs, their total (solid blue) expected contribution to the $\Omega_{\textrm{GW}}$ is shown. The dashed red, dot-dashed green and dotted orange lines show the contribution from each class of BBHs, with the relevant rates evaluated in \cite{TheLIGOScientific:2016pea}. 
\textit{Bottom}: assuming the SFR of \cite{Vangioni:2014axa} and negligible time delays, we show 
the expected $\Omega_{\textrm{GW}}$ for different assumptions on the mass distribution of BBHs.
If binaries with intermediate mass BHs larger than 100 $M_{\odot}$ exist, then the $\Omega_{\textrm{GW}}$ does not scale as $\propto f^{2/3}$ at frequencies $\gsim 10$ Hz. A (gray) line scaling at $\propto f^{2/3}$ is shown as a guide to the eye.}
\label{fig:StochGWBackMassAssup}
\end{figure}

Given that the three events detected, have very different masses, their classes contribute in the GW energy density at different frequency ranges. That is clearly shown in Figure~\ref{fig:StochGWBackMassAssup} (top panel), where we show the individual contribution 
from those classes of events and also their sum. 
The exact contributions, rely heavily on the rates. Yet, larger masses contribute more in lower 
frequencies. In fact, with future expected sensitivity, at low frequencies, we can probe the most massive members among the BBH population. \cite{Crowther:2010cg} has detected stars in the R136 cluster of the Large Magellanic Cloud that are more massive than 150 $M_{\odot}$; previously considered as unlikely to exist. 
In Figure~\ref{fig:StochGWBackMassAssup} (bottom), we show what the impact on $\Omega_{GW}$
would be if BHs with masses larger than 100 $M_{\odot}$ and up to 300 (or 500) $M_{\odot}$ exist and with the mass function still scaling as $\propto m_{1}^{-2.35}$. 
This is shown by comparing the blue solid vs the red dashed lines (or green dot-dashed lines). The 
important difference here is that if such massive binaries with at least one intermediate BH exist, then at frequencies between 10 and 50 Hz, 
a deviation from $\Omega_{GW}$ scaling as $f^{2/3}$ should be observable (see gray thin line that
shows $\Omega_{GW} \propto f^{2/3}$, given to guide the eye). 
We calculate a $S/N$ between 3.5 and 5 for LIGO which with ET-B design 
increases by a factor of 50-70 to values of 230-300 depending on the exact 
plotted $\Omega_{GW}$ spectrum.
Thus, the $\Omega_{GW}$ at those frequencies can be used to indirectly search for the most massive and rare stellar and the intermediate mass BHs in the Universe. This statement does not depend on any remaining uncertainties related to the energy released at the coalescence. 
Inversely, if very light BBH binaries merge at a high rate, their contribution would be seen at $f \sim O(10^{2})$ Hz (see orange dotted line in Figure~\ref{fig:StochGWBackMassAssup} bottom panel). 

\begin{figure}
\begin{centering}
\includegraphics[width=\columnwidth]{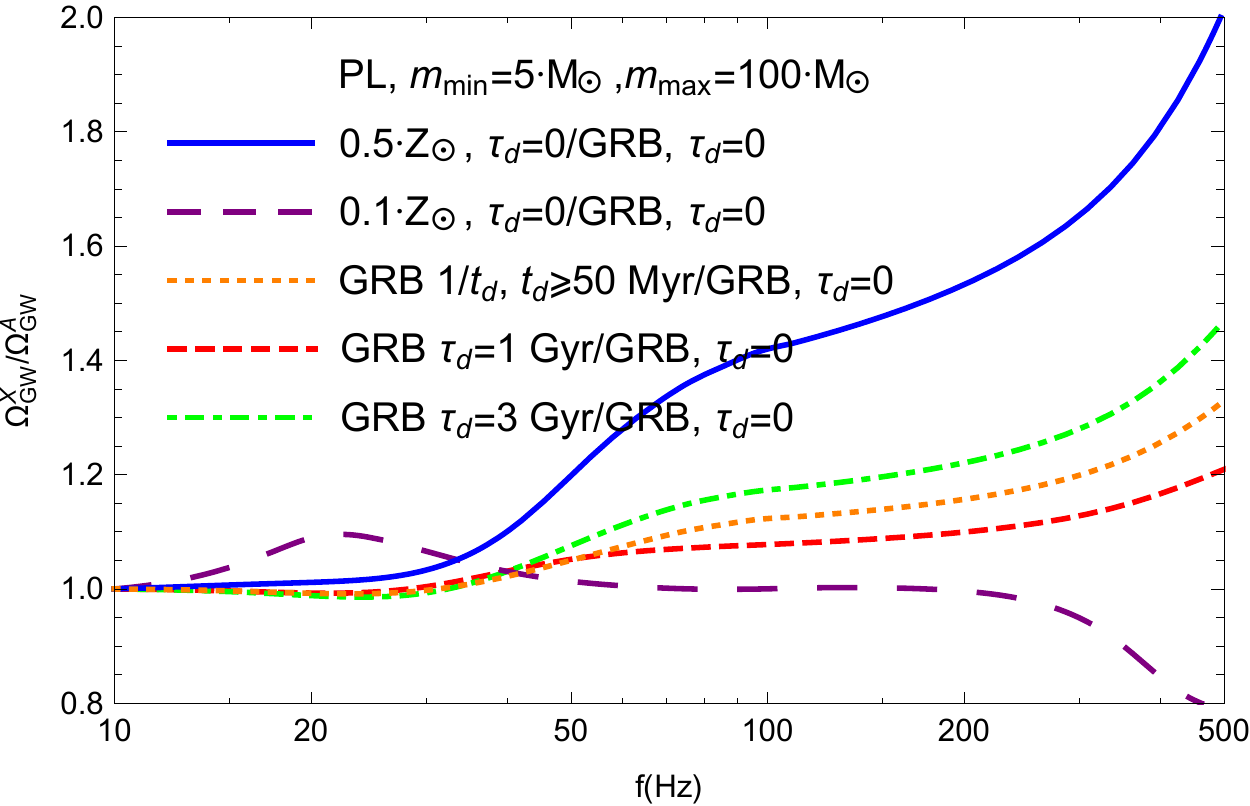}
\end{centering}
\caption{The ratio of $\Omega_{GW}^{X}/\Omega_{GW}^{A}$ where "X" and "A"
are different assumptions on the progenitors metallicity and time-delay. 
We assume fixed mass distribution for $m_{1}$ and $m_{2}$ of the BBHs. 
There is a strong degeneracy between the time delay, the environment of the BBH and the local merger rate, especially at $f_{\textrm{obs}} < 50$ Hz.}
\label{fig:StochGWBackRateAssump2}
\end{figure}

In Figure~\ref{fig:StochGWBackRateAssump2}, we show the impact that different assumptions on the
environment of formation of the stellar BBHs and on the time-delays, have on the $\Omega_{GW}$.
We calculate the ratio of $\Omega_{GW}^{X}/\Omega_{GW}^{A}$ between different set of assumptions "X" and "A". We always assume that the BBHs masses $m_{1}$ follow a PL distribution with $m_{\textrm{min}} = 5 M_{\odot}$
and $m_{\textrm{max}} = 100 M_{\odot}$, with a flat distribution on $q$.
As assumption "A", we always take the BBHs GRB redshift distribution with no
time delay. Assumption set "X" instead varies. We always normalize the ratio to 1 at 10 Hz.
There is a strong degeneracy 
between the redshift shape of $R_{m}(z)$ and its local normalization, at frequencies up to $\sim 50$ Hz relevant for LIGO.
That is shown more evidently by 
comparing reference assumptions "A" to the case where $R_{m}(z)$
has a distribution dominated by the environment with metallicities 
of 0.1 and 0.5 $Z_{\odot}$ respectively (see long dashed purple line and blue solid line and also Figure~\ref{fig:MergerRates}). 
With the ET that will measure $\Omega_{GW}$ up to $\simeq 300$ Hz some of those degeneracies will be addressed.
Moreover, the exact assumptions on the time-delay between binary formation and merger are strongly degenerate 
to the uncertainties in the local normalization of $R_{m}$. We show results with $P(t_{d}) \propto exp \{ -t_{d}/\tau_{d}\}$, with $\tau_{d}$ of 1 Gyr (red dashed line), 3 Gyr (green dot-dashed line) and with 
$P(t_{d}) \propto1/t_{d}$ with $t_{d} \ge 50$ Myr (orange dotted line). The gravitational wave energy density
alone will have little constraining power over those properties. Yet, we expect that the local rate of $R_{m}(z)$ will be well measured from the individual merger details by the end of the LIGO run at full design and
even better with the measurement from the ET \cite{Kovetz:2016kpi} ; allowing the decoupling of certain degeneracies. For the exact assumptions of Figure~\ref{fig:StochGWBackRateAssump2} we get a $S/N$ of 2.3-4 and 170-250 for LIGO and ET respectively.

\begin{figure}
\begin{centering}
\includegraphics[width=\columnwidth]{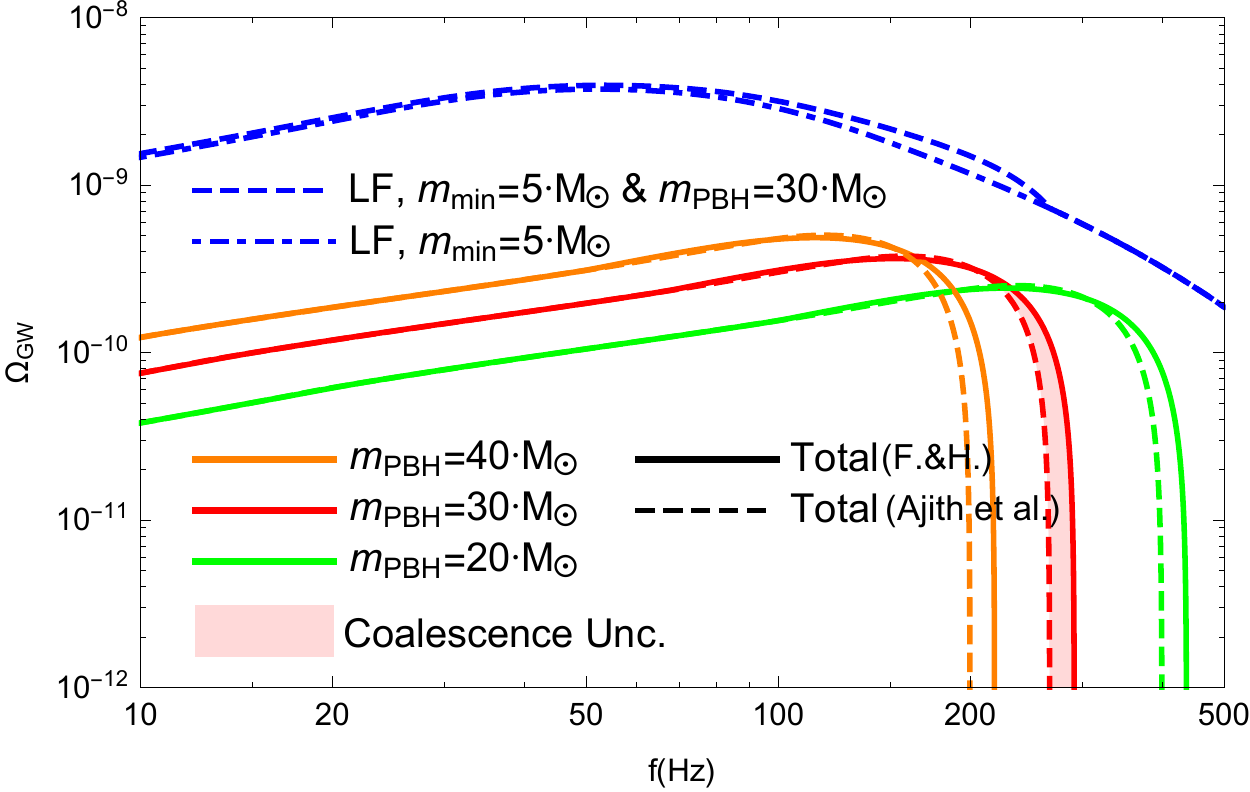}
\end{centering}
\caption{The contribution to the $\Omega_{GW}$ of a population of BBHs with a
monochromatic mass range based on \cite{Bird:2016dcv}, orange (red, green) for $m = 40$ (30, 20) $M_{\odot}$. In blue dot-dashed and dashed lines we show the contribution from stellar BBHs only and from stellar $\&$ primordial BH binaries respectively.
The uncertainties on the gravitational wave $\Omega_{GW}$ from the merger of PBHs, originating from the uncertainties in the mass of the PBHs and the exact frequency parametrization is shown as well.
The energy released during the last phases of the inspiral is fixed, while that of the merger and the ring-down phases differs between parameterizations (see text for details).}
\label{fig:StochPBHGWBack}
\end{figure}

An other possible probe for indirect searches of BBH properties through the $\Omega_{GW}$, is the search for features associated with the fact that specific models for the BBH progenitors may indicate 
a narrow mass range of the BH population. 
We calrify that measuring $\Omega_{GW}$ is going to be an additional tool to the measurement of the 
binaries masses and spins.
Since for a specific choice of masses there is a maximum 
frequency emitted at the quasi-normal ringdown $f_{\textrm{qnr}}$ (Eq.~\ref{eq:fqnr}); a significant population of BBHs with specific masses would contribute to $\Omega_{GW}$ with a specific frequency cut-off. 
That can result in possible spectral features on $\Omega_{GW}$.
In Figure~\ref{fig:StochPBHGWBack}, as an example, we show results for three different masses of PBHs.
If PBHs have a narrow mass range, then their relevant $\Omega_{GW}$ cut-off 
associated to the ring-down frequency may result in a bump in the total $\Omega_{GW}$. 
That can be seen in the specific plotted case between the 
blue dashed vs blue dot-dashed lines for an optimistic scenario. However, the 
merger rate of these PBHs remains small enough that we don't expect for either the GW residual
background or for the total $\Omega_{GW}$ to be within LIGO's reach \cite{TheLIGOScientific:2016wyq}.
For ET-B simplified assumptions, the difference in the $S/N$ between the two blue lines of Figure~\ref{fig:StochPBHGWBack}, is $\simeq 17$ ($S/N$ of 263 vs of 247).
We show results for 20, 30 and 40 $M_{\odot}$ PBHs (orange, red and green lines).
There are still some uncertainties in the physics of the emission of GWs during the coalescence.
We account for these, by performing calculations with both the parametrizations of \cite{Ajith:2009bn} 
and  \cite{Flanagan:1997sx} (dashed vs solid lines for the PBH component to the $\Omega_{GW}$).
The uncertainties are not large enough to diminish the importance of searching for features in the $\Omega_{GW}$, indicative of specific BBH population with future detectors.

\section{Conclusions}
\label{sec:Conclusions}

The detection of GWs from the merger event of binary BHs has opened a new window in astrophysics. In this work we discuss the importance of the gravitational waves energy density on probing the properties of the BBHs and thus possibly their origin. While the current uncertainties after only three merger events are 
still very wide, there is a series of questions that can be asked.

Since the GW energy density $\Omega_{GW}$ is the integrated merger rate of BBHs even at high redshifts where individual mergers can't be identified, its spectrum can give us information on the total BBH mass distribution.
We find that at high frequencies ($O(10^{2})$ Hz) the contribution from the lightest BHs can be probed 
and thus help us understand how often such objects form (see Figure~\ref{fig:StochGWBack} and Figure~\ref{fig:StochGWBackMassAssup}).

More interestingly, at frequencies as low as 10-50 Hz, we can indirectly search for a signal of the most massive BHs of stellar origin. If the initial stellar mass function extends to masses as large as $\sim 500 M_{\odot}$, 
resulting in BHs of $\gsim 100 M_{\odot}$ at the binaries, then a deviation from the expected 
$\Omega_{GW} \propto f^{2/3}$ spectrum behavior may be clearly observed with ET (see Figure~\ref{fig:StochGWBackMassAssup}).

The gravitational wave energy density amplitude depends strongly both on the total local rate and its redshift profile,
which in turn depends on the exact environment, the time of binary formation, and the time-delay between
formation and merger of the binary. The $\Omega_{GW}$ is strongly degenerate to those assumptions
(see Figure~\ref{fig:StochGWBackRateAssump2}). Yet, we can use the individual detected events 
from LIGO in the next years to measure the local value of $R_{m}$ and break some of these degeneracies.

Finally, we discussed that if populations of BHs exist with narrow mass distributions as are those of PBHs
\cite{Bird:2016dcv, Sasaki:2016jop, Clesse:2016vqa}, then mild spectral features may exist in the $\Omega_{GW}$, that future detectors 
can identify (see Figure~\ref{fig:StochPBHGWBack}).

The measurement of the spectrum of the gravitational wave energy density should be considered an other 
tool of searches to understand the properties of the BBHs in the Universe. That would be complementary 
to other studies as for instance cross-correlations of GW maps with galaxy catalogues \cite{Raccanelli:2016cud, Namikawa:2016edr}, searches for high modes of GW emission \cite{Seto:2016wom, Nishizawa:2016jji, Cholis:2016kqi, Nishizawa:2016eza} or
studies regarding the spins of the composite BHs \cite{Kalogera:1999tq, TheLIGOScientific:2016htt}.

\bigskip                  
                  
{\it Acknowledgements}: The author would like to thank Yacine Ali-Ha\"{i}moud, Ely Kovetz, Marc Kamionkowski, Chris Moore, Julian Mu\~{n}oz, Alvise Raccanelli and especially Simeon Bird and Vuk Mandic for interesting discussions. 
This work is supported by NASA Grant NNX15AB18G and the Simons Foundation.

\bibliography{GW_back}
\bibliographystyle{apsrev}

\end{document}